\documentclass[12pt]{iopart}

\usepackage[dvips]{epsfig}
\usepackage{amsfonts}
\usepackage{amssymb}
\usepackage{amstext}

\begin{document}

\title{Hierarchical Dobi\'nski-type relations via substitution and
the moment problem}
\author{K A Penson$^{\dag}$, P Blasiak$^{\dag\ddag}$, G Duchamp$^{\S\S}$, \\ A Horzela$^\ddag$
and A I Solomon$^{\dag\S}$}
\address{\ \linebreak$^{\dag}$Universit\'{e} Pierre et Marie Curie,\linebreak
Laboratoire   de  Physique   Th\'{e}orique  des  Liquides, CNRS UMR 7600\linebreak
Tour 16, $5^{i\grave{e}me}$ \'{e}tage, 4, place Jussieu, F 75252 Paris Cedex 05,
France\linebreak}
\address{$^{\ddag}$H. Niewodnicza{\'n}ski Institute of Nuclear Physics,
Polish Academy of Sciences\linebreak Department of Theoretical Physics\linebreak ul.
Radzikowskiego 152, PL 31-342 Krak{\'o}w, Poland\linebreak}
\address{$^\S$The Open University\linebreak
 Physics and Astronomy Department\linebreak
Milton Keynes MK7 6AA, United Kingdom\linebreak}
\address{$^{\S\S}$Universit\'{e} de Rouen\linebreak
Laboratoire d'Informatique Fondamentale et Appliqu\'{e}e de Rouen\linebreak 76821
Mont-Saint Aignan, France\linebreak} \eads{\linebreak\mailto{blasiak@lptl.jussieu.fr},
\mailto{gduchamp2@free.fr}, \mailto{andrzej.horzela@ifj.edu.pl},
\mailto{penson@lptl.jussieu.fr}, \mailto{a.i.solomon@open.ac.uk}\linebreak}

\begin{abstract}
We consider the transformation properties of integer sequences arising from the normal
ordering of exponentiated boson ($[a,a^\dag]=1$) monomials of the form $\exp[\lambda
(a^\dag)^ra]$, $r=1,2,\ldots$, under the composition of their exponential generating
functions (egf). They turn out to be of {\em Sheffer-type}. We demonstrate that two key
properties of these sequences remain preserved under substitutional composition: $a)$
the property of being the solution of the Stieltjes moment problem; and $b)$ the
representation of these sequences through infinite series (Dobi\'nski-type relations).
We present a number of examples of such composition satisfying properties $a)$ and
$b)$. We obtain new Dobi\'nski-type formulas and solve the associated moment problem
for several hierarchically defined combinatorial families of sequences.

\end{abstract}

\maketitle

\section{Introduction}

In a recent series of articles \cite{BPS1},\cite{BPS2},\cite{BPS3},\cite{BPS4},\cite{Krakow},\cite{Myczkowce} we investigated the properties of integer sequences
appearing in the process of the normal ordering of powers of boson monomials
$[(a^\dag)^r a^s]^n$, with $n$, $r$, $s$ -integers, where $a$ and $a^\dag$ are the
boson annihilation and creation operators respectively, satisfying $[a,a^\dag]=1$. They are extensions
of earlier works \cite{Katriel},\cite{Duchamp}. We observed that the normal form of
$[(a^\dag)^r a^s]^n$, with all the annihilation operators to the right, denoted by
${\mathcal N}\left([(a^\dag)^r a^s]^n\right)$, can be written  in the form ($r\geq
s$):
\begin{eqnarray}
[(a^\dag)^r a^s]^n \equiv {\mathcal N}\left([(a^\dag)^r
a^s]^n\right)=(a^\dag)^{n(r-s)}\sum_{k=s}^{ns}S_{r,s}(n,k)(a^\dag)^ka^k
\end{eqnarray}
where $S_{r,s}(n,k)$ are generalizations of the conventional ($r=s=1$) Stirling
numbers of the second kind and
\begin{eqnarray}
B_{r,s}(n)=\sum_{k=s}^{ns}S_{r,s}(n,k)
\end{eqnarray}
generalize the conventional ($r=s=1$) Bell numbers.

For general $r\geq s$ we have worked out a complete theory of the numbers
$S_{r,s}(n,k)$ and $B_{r,s}(n)$, including their recurrence relations, generating
functions and closed-form formulas. In particular, the generalized Bell numbers
$B_{r,s}(n)$ can be expressed as infinite series, thereby extending the celebrated
Dobi\'nski relation valid for $r=s=1$ \cite{Wilf}:
\begin{eqnarray}\label{Dobinski}
B_{1,1}(n)=\frac{1}{e}\sum_{k=0}^\infty\frac{k^n}{k!}, \ \ \ \ \ \ n=0,1,2,\ldots
\end{eqnarray}
Here are some examples of such relations:
\begin{eqnarray}
B_{r,1}(n)=\frac{1}{e}\sum_{k=1}^\infty\frac{1}{k!}\prod_{j=1}^n\left[k+(j-1)(r-1)\right]\\
B_{r,r}(n)=\frac{1}{e}\sum_{k=0}^\infty
\frac{1}{k!}\left[\frac{(k+r)!}{k!}\right]^{n-1}\label{Dobinski2}
\end{eqnarray}
they are all derived from the general polynomial-type formula ($n=1,2,\ldots$)
\begin{eqnarray}\nonumber
B_{r,s}(n,y)&=&\sum_{k=s}^{ns}S_{r,s}(n,k)y^k\\
&=&e^{-y}\sum_{k=s}^{\infty}\frac{1}{k!}
    \prod_{j=1}^n \left[\left(k+(j-1)(r-s)\right)
    \cdot\left(k+(j-1)(r-s)-1\right)\right.\cdot\nonumber\\
    &&\left. \ \ \ \ \ \ \ \ \ \ \ \ \ \ \ \ \ \ \ldots\cdot\left(k+(j-1)(r-s)-s+1\right)\right]y^k. \label{X}
\end{eqnarray}
We may associate a {\em Generating Function} $C(x)$ with a given
sequence $\{c_n\}$ by \cite{Wilf}
\begin{equation}
C(x)=\sum_{n=0}^{\infty}c_n{\frac{x^n}{n!}}.
 \end{equation}
 This particular form of Generating Function is known as a {\em Generating Function of
 Exponential Type} or {\em egf} for short, due to the $n!$ 
denominators.
 Of
particular interest for us here are those sequences $\{B_{r,s}(n)\}$ for which the
{\em egf} can in fact be expressed  as an exponential function; they include
$B_{r,1}(n)$, $r=1,2,...$ for which
\begin{eqnarray}\label{A}
e^{e^x-1}=\sum_{n=0}^\infty B_{1,1}(n)\frac{x^n}{n!}
\end{eqnarray}
and \cite{BPS1},\cite{BPS2},\cite{Lang}
\begin{eqnarray}\label{B}
\exp\left(\frac{1}{\sqrt[r-1]{1-(r-1)x^{r-1}}}-1\right)=\sum_{n=0}^\infty
B_{r,1}(n)\frac{x^n}{n!}.\ \ \ \ \ \ \ \ r=2,3,\ldots
\end{eqnarray}
The numbers $S_{r,s}(n,k)$ appear when in Eqs.(\ref{A}) and (\ref{B}) an indeterminate
$y$ is introduced through
\begin{eqnarray}\label{AA}
e^{y(e^x-1)}=\sum_{n=0}^\infty \left(\sum_{k=1}^n S_{1,1}(n,k)y^k\right)\frac{x^n}{n!}
\end{eqnarray}
and
\begin{eqnarray}\label{BB}
\exp\left[\textstyle y\left(\textstyle
\frac{1}{\sqrt[r-1]{1-(r-1)x^{r-1}}}-1\right)\right]=\sum_{n=0}^\infty
\left(\sum_{k=1}^n S_{r,1}(n,k)y^k\right)\frac{x^n}{n!},\\\nonumber&& r=2,3,\ldots
\end{eqnarray}
Eqs.(\ref{AA}) and (\ref{BB}) define  polynomials of order $n$:
\begin{eqnarray}\label{C}
B_{r,1}(n,y)=\sum_{k=1}^n S_{r,1}(n,k)y^k.\ \ \ \ \ \ \ \ r=1,2,\ldots.
\end{eqnarray}
Evidently, $B_{r,1}(n)=B_{r,1}(n,1)$. The polynomials of Eqs.(\ref{C}) share another
characteristic property: they can be written as ratios of two infinite series in $y$.
These are the so-called Dobi\'nski-type relations \cite{BPS1},\cite{BPS2}, which
 for $r=1$ and $r>1$ respectively are:
\begin{eqnarray}\label{D}
\frac{1}{e^y}\sum_{k=1}^\infty \frac{k^n}{k!}y^k=\sum_{k=1}^n S_{1,1}(n,k)y^k,\ \ \ \
\ \ \ n=0,1,\ldots
\end{eqnarray}
and
\begin{eqnarray}\label{E}
\frac{(r-1)^n}{e^y}\sum_{k=1}^\infty \frac{\Gamma (n+\frac{k}{r-1})}{k!\Gamma
(\frac{k}{r-1})}y^k=\sum_{k=1}^n S_{r,1}(n,k)y^k,\ \ \ \ \ \ \ n=1,2,\ldots.
\end{eqnarray}
By setting $y=1$ in Eqs.(\ref{AA}) and (\ref{BB}) we obtain a representation of the
integers  $B_{r,1}(n)$ as an infinite series (compare
Eqs.(\ref{Dobinski})-(\ref{Dobinski2})); this constitutes a fertile ground for their
probabilistic interpretation \cite{Pitman},\cite{Constantine}. The numbers
$B_{r,1}(n)$ can also be given various combinatorial intepretations
\cite{BDHPS-tbp},\cite{MPBS-tbp}. The second consequence of Eqs.(\ref{D}) and
(\ref{E}) (and of the more general formulas for $s>1$, see \cite{BPS1},\cite{BPS2} )
is the fact, that $B_{r,1}(n,y)$ for $y>0$ is the $n$-th Stieltjes moment of a
non-negative probability distribution, which is either discrete (for $r=1$, giving a
so called {\em Dirac comb} \cite{BPS3}) or continuous (for $r>1)$. This fact permits
one to use the $B_{r,s}(n,y)$ to construct various  quantum collective states called
coherent states \cite{BPS4},\cite{Klauder}. The interpretation of combinatorial sequences as
moments \cite{Krakow} has led to new calculational approaches to hyperdeterminants
\cite{Thibon}. Another aspect of Eqs.(\ref{D}) and (\ref{E}) which deserves mention
here is that the numbers $S_{r,1}(n,k)$ ($1\leq k\leq n$) form a non-singular
lower-triangular matrix with ones on the diagonal. Such matrices form a group, called
the {\em Riordan group}, which has important applications in  enumerative combinatorics
\cite{Shapiro},\cite{Chin}.

The purpose of this note is to place Eqs.(\ref{AA})-(\ref{E}) in the more general
context of {\em Sheffer-type} polynomials and to address the question of compositional
substitution and its implication for the existence of Dobi\'nski-type relations as
solutions of the Stieltjes moment problem.

We  first recall the known fact \cite{Stanley},\cite{Aldrovandi},\cite{Flajolet} that
a compositional substitution corresponds to multiplication of the matrices
$S_{r,1}(n,k)$. Then we go on to demonstrate that if two polynomial sequences
$B_F(n,y)$ and $B_G(n,y)$ generated by $e^{yF(x)}$ and $e^{yG(x)}$ respectively are
solutions of the associated Stieltjes moment problems, then the sequence
$B_{F(G)}(n,y)$ is also a solution of another, closely related,  Stieltjes moment
problem. We further prove that if $B_F(n,y)$ and $B_G(n,y)$ are both given by
Dobi\'nski-type relations, see Eqs.(\ref{D}) and (\ref{E}), then the sequence
$B_{F(G)}(n,y)$ is also given by an analogous formula. We then illustrate these {\it
reproducing} properties of Dobi\'nski-type relations and  moment problem solutions by
some specific examples. They comprise multiple compositions of standard Bell numbers
with themselves (composing discrete with discrete distributions), compositions of Lah
numbers (related to Laguerre polynomials) with themselves and finally composing
discrete with continuous distributions and vice versa.

\section{Sheffer-type polynomials}

A polynomial $B(n,y)$ of order $n$ in the variable $y$ is of Sheffer-type if the
associated  egf can be written in the form \cite{Roman}
\begin{eqnarray}\label{H}
1+\sum_{n=1}^\infty B(n,y)\frac{x^n}{n!}=A(x)e^{yF(x)}
\end{eqnarray}
with $A(0)=1$ and $F(0)=0$. Many such families of polynomials have been thoroughly
investigated. Among the polynomials encountered in Quantum Mechanics, the Hermite and
Laguerre polynomials are of Sheffer-type, whereas the Legendre and Gegenbauer are not.
Comparing Eq.(\ref{H}) with Eqs.(\ref{AA}) and (\ref{BB}) we observe that
$B_{r,1}(n,y)$ are Sheffer-type polynomials with $A(x)=1$. In fact $B_{1,1}(n,y)$ are
the so-called Bell (or exponential) polynomials \cite{Roman} and $B_{2,1}(n,y)$ are
the generalized Laguerre polynomials. The numbers $S_{1,1}(n,k)$ are the conventional
Stirling numbers of the second kind and the numbers
\begin{eqnarray}\label{Lah}
S_{2,1}(n,k)=\frac{n!}{k!}{n-1\choose k-1}
\end{eqnarray}
are the so-called unsigned Lah numbers \cite{BPS1},\cite{Lang}.

More generally, consider two families of Sheffer-type polynomials $B_F(n,y)$ and
$B_G(n,y)$ generated by
\begin{eqnarray}\label{exp(yF)}
e^{yF(x)}=1+\sum_{n=1}^\infty \left(\sum_{k=1}^nS_F(n,k)y^k\right)\frac{x^n}{n!}
\end{eqnarray}
and
\begin{eqnarray}\label{exp(yG)}
e^{yG(x)}=1+\sum_{n=1}^\infty \left(\sum_{k=1}^nS_G(n,k)y^k\right)\frac{x^n}{n!},
\end{eqnarray}
respectively, where $F(0)=G(0)=0$ and
\begin{eqnarray}
B_F(n,y)=\sum_{k=1}^nS_F(n,k)y^k\ \ \ \ \text{and}\ \ \ \
B_G(n,y)=\sum_{k=1}^nS_G(n,k)y^k.
\end{eqnarray}
We now consider  the polynomials generated by $F(G(x))$, i.e.
\begin{eqnarray}\label{exp(FG)}
e^{yF(G(x))}=1+\sum_{n=1}^\infty
\left(\sum_{k=1}^nS_{F(G)}(n,k)y^k\right)\frac{x^n}{n!}
\end{eqnarray}
Before we calculate this sum we note the relation resulting from the change of
summation in Eq.(\ref{exp(yG)}):
\begin{eqnarray}\label{EE}
e^{yG(x)}=1+\sum_{k=1}^\infty \left(\sum_{n=k}^\infty S_G(n,k)\frac{x^n}{n!}\right)y^k.
\end{eqnarray}
Now comparison with the direct expansion of the left hand side of Eq.(\ref{EE})
\begin{eqnarray}
e^{yG(x)}=1+\sum_{k=1}^\infty (G(x))^k\ y^k/ k!
\end{eqnarray}
yields
\begin{eqnarray}\label{G(x)^n}
\frac{(G(x))^k}{k!}=\sum_{n=k}^\infty S_G(n,k)\frac{x^n}{n!}.
\end{eqnarray}
Proceeding to the direct calculation of Eq.(\ref{exp(FG)}) we recall
 that the matrices $S_F(n,k)$ and $S_G(n,k)$ are lower
triangular (i.e. the entries for $k>n$ are zero):
\begin{eqnarray}
e^{yF(G(x))}&=&1+\sum_{n=1}^\infty
\left(\sum_{k=1}^nS_F(n,k)y^k\right)\frac{(G(x))^n}{n!}\\
&=&1+\sum_{n=1}^\infty
\left(\sum_{k=1}^nS_F(n,k)y^k\right)\sum_{p=n}^\infty S_G(p,n)\frac{x^p}{p!}\\
&=&1+\sum_{p=1}^\infty
\left(\sum_{k=1}^p\left(\sum_{n=1}^pS_G(p,n)S_F(n,k)\right)y^k\right)\frac{x^p}{p!}.
\end{eqnarray}
Comparison with Eq.(\ref{exp(FG)}) yields
\begin{eqnarray}\label{I}
S_{F(G)}(n,k)=\sum_{p=1}^nS_G(n,p)S_F(p,k)
\end{eqnarray}
This last equality means that  compositional substitution within the Sheffer-type
polynomial families is equivalent to the matrix product of the corresponding Stirling
matrices \cite{Stanley},\cite{Aldrovandi}
\begin{eqnarray}
\mathbb{S}_{F(G)}=\mathbb{S}_G\cdot\mathbb{S}_F\ \ .
\end{eqnarray}
A direct consequence of Eq.(\ref{I}) is the formula
\begin{eqnarray}
B_{F(G)}(n,y)&=&\sum_{p=1}^nS_{F(G)}(n,p)y^p=\sum_{p=1}^ny^p\sum_{k=1}^nS_G(n,k)S_F(k,p)\\
&=&\sum_{k=1}^nS_G(n,k)\sum_{p=1}^kS_F(k,p)y^p=\sum_{k=1}^nS_G(n,k)B_F(k,y).\label{J}
\end{eqnarray}
The last equation can be seen as the generalized Stirling transform
\cite{SloaneStirling} of the polynomials $B_F(k,y)$ which for $y=1$ reduces to the
generalized Stirling transform of the sequence $B_F(k)$:
\begin{eqnarray}\label{JJ}
B_{F(G)}(n)=\sum_{k=1}^nS_G(n,k)B_F(k).
\end{eqnarray}

\section{Compositional moment problem}

The formulas (\ref{I}) and (\ref{J}) lead to important consequences if the initial
Sheffer-type polynomials are solutions of the Stieltjes moment problems, i.e. if for
$x,y>0$ there exist positive weight functions $W_F(x,y)$ and $W_G(x,y)$ such that
\begin{eqnarray}
B_F(n,y)=\int_0^\infty x^nW_F(x,y)dx,\\
B_G(n,y)=\int_0^\infty x^nW_G(x,y)dx\ .
\end{eqnarray}
Then the following equalities follow:
\begin{eqnarray}\nonumber
B_{F(G)}(n,y)=\sum_{k=1}^nS_G(n,k)B_F(k,y)\\\nonumber=\sum_{k=1}^nS_G(n,k)\int_0^\infty
x^kW_F(x,y)dx =\int_0^\infty W_F(x,y)\sum_{k=1}^nS_G(n,k)x^k\ dx\\
\nonumber=\int_0^\infty W_F(x,y)B_G(n,x)dx =\int_0^\infty dx\ W_F(x,y)\int_0^\infty
z^n W_G(z,x)dz\\=\int_0^\infty z^n \left(\int_0^\infty W_F(x,y)W_G(z,x)dx\right)dz
\end{eqnarray}
and this implies that
\begin{eqnarray}
B_{F(G)}(n,y)=\int_0^\infty x^nW_{F(G)}(x,y)dx
\end{eqnarray}
where $W_{F(G)}(x,y)$ is a positive function given by
\begin{eqnarray}\label{Weight}
W_{F(G)}(x,y)=\int_0^\infty W_F(z,y)W_G(x,z)dz.
\end{eqnarray}
We remark that the arguments of the weight functions in Eq.(\ref{Weight}) need not
satisfy any particular symmetry properties.

More generally, for $p$-fold substitution $F_1(F_2(\ldots(F_p)\ldots))$ one obtains
\begin{eqnarray}\nonumber
W_{F_1(F_2(\ldots(F_p)\ldots))}(x,y)&=&\int_0^\infty dz_1\ W_{F_1}(z_1,y)\int_0^\infty
dz_2\ W_{F_2}(z_2,z_1)\ldots\\&&\ \ \ \ \ \  \ldots \int_0^\infty dz_p\
W_{F_{p-1}}(z_p,z_{p-1})W_{F_p}(x,z_p).\label{FF}
\end{eqnarray}
Eq.(\ref{FF}) reveals a typical structure appearing in the iterated-kernel method of
solving integral equations \cite{Morse},\cite{Krasnov}.

In other words; for the Sheffer-type polynomials the property of being a solution of
the Stieltjes moment problem is {\it reproduced} by the mechanism of compositional
substitution, under the evident condition that the integrals in Eqs.(\ref{Weight}) and
(\ref{FF}) exist. In the following section we provide  a number of examples of
substitutions $F(G(x))$ for which an explicit evaluation of $W_{F(G)}(x,y)$ and
$B_{F(G)}(n,y)$ can be carried through.

\section{Compositional Dobi\'nski-type relations}

A rather large reservoir of solutions of the Stieltjes moment problem is contained in
the formulas (\ref{D}) and (\ref{E}). For any $r=1,2,\ldots$ $B_{r,1}(n,y)$ is the
moment of a positive function $W_r(x,y)$, which can be written down explicitly, for
instance by extending to $y\neq1$ the results given in \cite{BPS4},\cite{Krakow},
\cite{Myczkowce}. The examples are:
\begin{eqnarray}
W_1(x,y)&=&e^{-y}\sum_{k=1}^\infty \frac{y^k\delta(x-k)}{k!}\\\label{W1}
W_2(x,y)&=&ye^{-(x+y)}\frac{I_1(2\sqrt{xy})}{\sqrt{xy}}\\\nonumber
W_3(x,y)&=&\frac{1}{12\sqrt{\pi}x}e^{-\frac{x}{2}-y}y\left(6\sqrt{2x}+3xy\sqrt{\pi}\
{_0F_2}(\textstyle\frac{3}{2},2;\textstyle\frac{xy^2}{8})\right.\\&&\ \ \ \ \ \ \ \ \ \ \ \ \ \ \ \ \ \ \ \ \ \ \ 
\left.+\sqrt{2}x^{3/2}y^2\
{_1F_3}(1;\textstyle\frac{3}{2},2,\textstyle\frac{5}{2};\textstyle\frac{xy^2}{8})\right),\label{W2}
\end{eqnarray}
where $\delta(z)$ is the Dirac delta function, $I_\nu (z)$ is the modified Bessel function of first kind and $_0F_2$ and
$_1F_3$ are hypergeometric functions. Eqs.(\ref{W1}) and (\ref{W2}) were obtained
using the inverse Mellin transform. See \cite{PS} for its exposition and \cite{Quesne}
for examples of applications.

Note, that whereas $W_1(x,y)$ is a discrete distribution in the  form of a  Dirac comb
concentrated on positive integers, the functions $W_r(x,y)$ for $r>1$ are continuous
distributions \cite{Myczkowce}. Observe also that they are  not normalized, in the
sense of their zero moments: $\int_0^\infty W_1(x,1)dx=1$ whereas $\int_0^\infty
W_r(x,1)dx\neq 1$, $r>1$.

In this section we demonstrate that the reproducing character of the compositional
moment problem, see Eq.(\ref{Weight}), implies the reproducing character of the
Dobi\'nski-type relations. In the following paragraph, with given $F(x)$ and $G(x)$ of
Eqs.(\ref{exp(yF)}) and (\ref{exp(yG)}) we will carry out explicit substitutions
$F(G(x))$ and analyze the weight functions $W_{F(G)}(x)$ obtained from
Eq.(\ref{Weight}) and the resulting Dobi\'nski-type relations.

\subsection{$F(x)=G(x)=e^x-1$}\label{11}

In the following the subscript $B(B)$ stands for  ``substitute Bell into Bell''. We
investigate the polynomials $B_{B(B)}(n,y)$ resulting from
\begin{eqnarray}
e^{\textstyle y\left(e^{e^x-1}-1\right)}=\sum_{n=0}^\infty B_{B(B)}(n,y)\frac{x^n}{n!}
\end{eqnarray}
which correspond to the ordinary Stirling transform \cite{SloaneStirling} of the Bell
polynomials $B_{1,1}(n,y)$
\begin{eqnarray}
B_{B(B)}(n,y)=\sum_{k=1}^nS(n,k)B_{1,1}(k,y)
\end{eqnarray}
where $S(n,k)$ are the conventional Stirling numbers of the second kind. The
polynomial $B_{1,1}(n,y)$ is the $n$-th moment of the Dirac comb \cite{BPS3},
\begin{eqnarray}\label{YY1}
W_B(x,y)=e^{-y}\sum_{k=1}^\infty \frac{y^k\delta(x-k)}{k!},
\end{eqnarray}
and the weight function resulting from the substitution $F(F(x))$ is through
Eq.(\ref{Weight}) equal to
\begin{eqnarray}\nonumber
W_{B(B)}(x,y)&=&\int_0^\infty W_B(z,y)W_B(x,z)dz=\\\nonumber
&=&\int_0^\infty\left(e^{-y}\sum_{k=1}^\infty
\frac{y^k\delta(z-k)}{k!}\right)\left(e^{-z}\sum_{p=1}^\infty
\frac{z^p\delta(x-p)}{p!}\right)dz\\\nonumber &=&e^{-y}\sum_{p=1}^\infty
\frac{\delta(x-p)}{p!}\left(\sum_{k=1}^\infty
\frac{k^p}{k!}(ye^{-1})^k\right)\\
&=&e^{y(e^{-1}-1)}\sum_{p=1}^\infty
\frac{\delta(x-p)}{p!}\left(\sum_{r=1}^pS(p,r)(ye^{-1})^r\right),
\end{eqnarray}
where the last equality results from the original Dobi\'nski formula Eq.(\ref{D}).
This result shows that
\begin{eqnarray}\label{YY2}
B_{B(B)}(n)=B_{B(B)}(n,1)=e^{\textstyle(e^{-1}-1)}\sum_{k=1}^\infty
\frac{k^n}{k!}\left(\sum_{r=1}^p S(k,r)e^{-r}\right),
\end{eqnarray}
with the initial terms $B_{B(B)}(n)=1, 1,3,12,60,358,2471,19302,\ldots$, for
$n=0,1,\ldots$ . $B_{B(B)}(n)$ counts the number of partitions of a set of $n$
distinguishable elements, in which every part is again partitioned \cite{Stanley}.

Multiple substitutions of Bell egf's into themselves result in hierarchical,
chain-like formulas for corresponding partition numbers, i.e. for $F(F(F(x)))$ one
obtains for $n=0,1,\ldots$
\begin{eqnarray}\nonumber
B_{B(B(B))}(n)&=&\e^{\textstyle (e^{ (e^{-1}-1)}-1)}\sum_{k=1}^\infty
\frac{k^n}{k!}\cdot \\\label{YY3}
&&\ \ \ \cdot \left(\sum_{p=1}^kS(k,p)e^{-p}\left(\sum_{r=1}^p
S(p,r)e^{r(e^{-1}-1)}\right)\right).
\end{eqnarray}
For example, $B_{B(B(B))}(n)=1,1,4,22,154,1304,12915,146115,\ldots$, for
$n=0,1\ldots$, which counts the number of ``triple'' partitions of an $n$-set.

We conclude that the substitution $F(F(x))$ results in a formula for $B_{B(B)}(n)$
which conserves the original Dobi\'nski-type structure of $B_B(n)$ as in
Eq.(\ref{Dobinski}); and also gives  a Dirac comb type of weight function  with
modified weights concentrated on positive integers. These results also hold good for
higher order substitutions.

\subsection{$F(x)=G(x)=\frac{x}{1-x}$}\label{22}

This case corresponds to $B_{2,1}(n,y)$  which  from Eq.(\ref{E})  is
\begin{eqnarray}
B_{2,1}(n,y)=\frac{1}{e^y}\sum_{k=1}^\infty \frac{\Gamma(n+k)}{k!\Gamma
(k)}y^k=n!\sum_{k=1}^n\frac{1}{k!}{n-1\choose k-1}y^k,
\end{eqnarray}
and can be also written as
\begin{eqnarray}\label{L}
B_{2,1}(n,y)=(n-1)!\ yL_{n-1}^{(1)}(-y)
\end{eqnarray}
by using the standard form of the generating function of generalized Laguerre
polynomials $L_n^{(\lambda)}(x)$. With the notational convention introduced  above we
rewrite Eq.(\ref{L}) as (here $L$ stands for Laguerre)
\begin{eqnarray}
B_L(n,y)=\sum_{k=1}^nS_L(n,k)y^k
\end{eqnarray}
where $S_L(n,k)$ are the unsigned Lah numbers, see Eq.(\ref{Lah}). For $y=1$, the
integers $B_L(n,1)\equiv B_L(n)$ count binary ordered forests of $n$ nodes
\cite{BDHPS-tbp} (the initial terms are $B_L(n)=1,3,13,73,501,4051\ldots$ ,
$n=1,2,\ldots$). For other combinatorial interpretations see \cite{EIS}.

The polynomial $B_L(n,y)$ is the $n$-th moment of \cite{Myczkowce}
(see Eq.(\ref{W2}) ):
\begin{eqnarray}\label{WL}
W_L(x,y)=ye^{-(x+y)}\frac{I_1(2\sqrt{xy})}{\sqrt{xy}}
\end{eqnarray}
By  $F(F(x))$-type composition the function $\exp\left(\frac{yx}{1-2x}\right)$
generates $B_{L(L)}(n,y)$ through
\begin{eqnarray}
e^{ y\frac{x}{1-2x}}=\sum_{n=0}^\infty B_{L(L)}(n,y)\frac{x^n}{n!}
\end{eqnarray}
where $L(L)$ stands for  ``substitute Laguerre into Laguerre'', which are the $n$-th
moments of
\begin{eqnarray}\nonumber
W_{L(L)}(x,y)&=&\int_0^\infty W_L(z,y)W_B(x,z)dz=\\
&=&\int_0^\infty\left(ye^{-(z+y)}\frac{I_1(2\sqrt{zy})}{\sqrt{zy}}\right)\cdot
\left(ze^{-(x+z)}\frac{I_1(2\sqrt{xz})}{\sqrt{xz}}\right)dz.
\end{eqnarray}
By virtue of the entry $2.15.20.8$ of \cite{Brych}, this yields a continuous
distribution
\begin{eqnarray}
W_{L(L)}(x,y)=ye^{-\frac{x+y}{2}}\frac{I_1(\sqrt{xy})}{2\sqrt{xy}}=\frac{1}{2}W_L(\textstyle\frac{x}{2},\textstyle\frac{y}{2}),
\end{eqnarray}
thus preserving the original structure encountered in Eq.(\ref{WL}). In addition,
simple use of the generating function of the generalized Laguerre polynomials yields
\begin{eqnarray}\label{1}
B_{L(L)}(n,y)=\int_0^\infty x^n W_{L(L)}(x,y)dx=2^{n-1}(n-1)!\
yL_{n-1}^{(1)}(\textstyle-\frac{y}{2})
\end{eqnarray}
whose initial terms for $y=1$ are $B_{L(L)}(n)=1,5,37,361,4361,62701\ldots$,
$n=1,2,\ldots$ . The $p$-fold substitution, $p=1,2,\ldots$ , gives in this case the
compact expression:
\begin{eqnarray}\label{2}
B_{L(L(\ldots(L)\ldots))}(n)=p^{n-1}(n-1)!L_{n-1}^{(1)}(\textstyle-\frac{1}{p}),\ \ \
\ \ \ n=1,2,\ldots
\end{eqnarray}

\subsection{$F(x)=e^x-1,\ G(x)=\frac{x}{1-x}$}

Here we substitute Laguerre (continuous distribution) into Bell (discrete
distribution) and vice versa.

The calculations are analogous to those in \ref{11} and \ref{22} with repeated use of
integrals listed in \cite{Brych}. We  only quote the final results:
\begin{eqnarray}
B_{B(L)}(n,y)=\int_0^\infty x^n W_{B(L)}(x,y)dx,
\end{eqnarray}
where
\begin{eqnarray}
W_{B(L)}(x,y)=\frac{e^{-(x+y)}}{\sqrt{x}}\sum_{k=1}^\infty
\frac{y^k}{k!}\sqrt{k}e^{-k}I_1(2\sqrt{kx})
\end{eqnarray}
which is a continuous distribution. The polynomials $B_{B(L)}(n,y)$ are generated by
\begin{eqnarray}
e^{ y(e^{\frac{x}{1-x}}-1)}=\sum_{n=0}^\infty B_{B(L)}(n,y)\frac{x^n}{n!}.
\end{eqnarray}
The initial terms of $B_{B(L)}(n)$ are $1,4,23,171,1552,16583\ldots$ for $n=1,2,\ldots$. These integers
count structures called {\em sets of sets of lists}, where {\em list} means an ordered
subset \cite{EIS}. A closed-form Dobi\'nski-type formula for $B_{B(L)}(n)$ can be
obtained by calculating the moments of $W_{B(L)}(x,1)$. A longer but straightforward
calculation gives
\begin{eqnarray}\label{3}
B_{B(L)}(n)=e^{-1}\sum_{k=1}^\infty \frac{(n-1)!L_{n-1}^{(1)}(-k)}{(k-1)!}\ .
\end{eqnarray}
Higher order substitutions yield formulas of similar type.

For the opposite substitution (``Bell into Laguerre'' denoted $L(B)$ ) generated by
\begin{eqnarray}
e^{ y\textstyle\frac{e^x-1}{2-e^x}}=\sum_{n=0}^\infty B_{L(B)}(n,y)\frac{x^n}{n!}.
\end{eqnarray}
we obtain
\begin{eqnarray}
B_{L(B)}(n,y)=\int_0^\infty x^n W_{L(B)}(x,y)dx,
\end{eqnarray}
where
\begin{eqnarray}
W_{L(B)}(x,y)=\frac{y}{2}e^{-\frac{y}{2}}\sum_{k=1}^\infty \frac{\delta(x-k)}{2^k\cdot
k}L_{k-1}^{(1)}(\textstyle-\frac{y}{2})
\end{eqnarray}
which is a discrete (Dirac comb) distribution, with moments
\begin{eqnarray}\label{4}
B_{L(B)}(n)=B_{L(B)}(n,1)=\frac{1}{2}e^{-\frac{1}{2}}\sum_{k=1}^\infty
\frac{k^{n-1}}{2^k}L_{k-1}^{(1)}(\textstyle-\frac{1}{2})
\end{eqnarray}
and initial terms $B_{B(L)}(n)=1,4,23,173,1602,17575\ldots$, for $n=1,2,\ldots$ .

\subsection{Bell numbers vs. ``ordered'' Bell numbers}

As the last example we shall consider a slightly more general substitution problem in
which only the ``internal'' egf  $G(x)$  is of Sheffer-type. In
other words, the egf of one of the sequences is not an exponential. A case in point is
given by the so called ``ordered'' Bell numbers \cite{BPS3},\cite{Wilf} $B_O(n)$
defined through
\begin{eqnarray}
B_O(n)=\sum_{k=1}^n S(n,k)\ k!\ .
\end{eqnarray}
Their extension to polynomials $B_O(n,y)=\sum_{k=1}^nS(n,k)\ k!\ y^k$ is generated by
\cite{Flajolet}
\begin{eqnarray}
\frac{1}{1-y(e^x-1)}=\sum_{n=0}^\infty B_O(n,y)\frac{x^n}{n!}.
\end{eqnarray}
Thus the $B_O(n,y)$ are not of Sheffer-type.

We now perform the substitution ``Bell into ordered Bell'', denoted by the subscript
$O(B)$.  Although  Eq.(\ref{J}) is no longer valid, we can still define the numbers
$B_{O(B)}(n)$ through Eq.(\ref{JJ}):
\begin{eqnarray}
B_{O(B)}(n)=\sum_{k=1}^nS(n,k)B_O(k),
\end{eqnarray}
or equivalently by
\begin{eqnarray}
\frac{1}{2-e^{e^x-1}}=\sum_{n=0}^\infty B_{O(B)}(n)\frac{x^n}{n!}.
\end{eqnarray}
Recalling the Dobi\'nski-type expression for $B_O(n)$ \cite{BPS3}, \cite{Wilf}
\begin{eqnarray}\label{YY4}
B_O(n)=\frac{1}{2}\sum_{k=0}^\infty \frac{k^n}{2^k}\; ,
\end{eqnarray}
the formula Eq.(\ref{Weight}), now for $y=1$ only, carries over and after
straightforward calculation we obtain the Dobi\'nski-type formula for $B_{O(B)}(n)$:
\begin{eqnarray}\label{OB}\label{5}
B_{O(B)}(n)=\frac{1}{2}\sum_{k=0}^\infty \frac{k^n}{k!}Li_{-k}(\textstyle\frac{1}{2e})
\end{eqnarray}
where $Li_m(y)$ is the polylogarithm of order $m$ of $y$. The initial terms are
$B_{O(B)}(n)=1,4,23,175,1662,18937,\ldots$, $n=1,2,\ldots$ .

Similarly, from the  substitution ``double Bell into ordered Bell'' (denoted by
$O(B(B))$ below) we obtain
\begin{eqnarray}\label{OBB}
B_{O(B(B))}(n)=\frac{1}{2}\sum_{k=0}^\infty
\frac{k^n}{k!}\left(\sum_{r=1}^\infty\frac{e^{-r}r^k}{r!}
Li_{-r}(\textstyle\frac{1}{2e})\right)
\end{eqnarray}
where
\begin{eqnarray}
\frac{1}{2-e^{e^{e^x-1}-1}}=\sum_{n=0}^\infty B_{O(B(B))}(n)\frac{x^n}{n!}\ ,\ \ \ \ \
\ \ \ etc.
\end{eqnarray}
Clearly, Eqs.(\ref{OB}) and (\ref{OBB}) again give rise to  Dirac comb weight
functions.

\section{Discussion and conclusions}

The main result of this work can be viewed from different perspectives. It is
primarily a method for the  generation of new solutions of moment problems. As such it
is of potential importance for  the construction of new generalized coherent states.
Refs. \cite{Krakow} and \cite{BPS4} should be considered as first steps in this
direction. The iterative method based on Eqs.(\ref{Weight}) and (\ref{FF}) appears to
be straightforward under the condition of the existence of the relevant integrals.
This will definitely extend and enrich the families of currently known solutions of
the moment problem.

A closer look at the examples above based on  Eq.(\ref{Weight})
leads to the conclusion that if $e^{G(x)}$ generates the moments of a discrete distribution then the
moments generated by $e^{F(G(x))}$ 
are  those of  a discrete
distribution. Similarly, when  $e^{G(x)}$ gives a 
continuous distribution, the composition $e^{F(G(x))}$ gives rise to a 
continuous distribution.

We are dealing here with Sheffer-type polynomials which are also  solutions of the
moment problem; it should be borne in mind that these are quite strong restrictions.
It is easy to construct Sheffer-type polynomials which are {\em not} solutions of the
moment problem. For example, the polynomials $p_n(y)$, which are
related to Bessel polynomials \cite{Roman}, are generated by
\begin{eqnarray}
e^{y(\sqrt{1+2x}-1)}=1+\sum_{n=1}^\infty p_n(y) \frac{x^n}{n!}
\end{eqnarray}
and can take on negative values for $y=1$; they   are  therefore  not acceptable solutions of the moment
problem. On the other hand, for $s>1$, the polynomials  $B_{r,s}(n,y)$ defined by
Eq.(\ref{X}) are solutions of the moment problem \cite{Myczkowce} but
are {\em not} of
Sheffer-type \cite{BPS1},\cite{BPS2}.

Referring to various Dirac comb-type distributions obtained by
compositions (see Eqs.(\ref{YY1}), (\ref{YY2}), (\ref{YY3}),
(\ref{4}), (\ref{YY4}), (\ref{5}) and (\ref{OBB}) ) we
observe that the substitution  $B(n)\to B(\alpha n^2+\beta n+\gamma)$, ($\alpha,\beta,\gamma$ -
integers, $\alpha>0$) gives sequences $\tilde{B}(n)=B(\alpha n^2+\beta n+\gamma)$ which are the $n$-th moments of
continuous measures; they are infinite, weighted sums of
log-normal distributions \cite{BPS3}.

The reproducing nature of Dobi\'nski-type relations under composition also follows
from the scheme presented here. It has already provided a number of new closed-form
expressions for combinatorial numbers,
Eqs.(\ref{1}),(\ref{2}),(\ref{3}),(\ref{4}),(\ref{5}) and (\ref{OBB}), together with
the associated weight functions. It seems that this method can be also applied to
various generalizations of combinatorial numbers, e.g. $q$-deformations \cite{Schork}
and to more involved substitution schemes such as those considered in \cite{Sloane}.

\ack

We thank D. Barsky, M. M. Mendez and C. Quesne for important discussions.

\end{document}